\documentclass[english,12pt,twoside]{article}
\usepackage{amssymb}
\usepackage{amsmath}
\usepackage{babel}
\usepackage{array}
\usepackage[dvips]{graphicx}
\usepackage{pstricks}
\usepackage{pst-node}
\input{Qcircuit}

\date{}
\title{Quantum Games and Programmable Quantum Systems}
\author{
Edward W. Piotrowski\\ Institute of Mathematics, University of
Bia\l ystok,\\ Lipowa 41, Pl 15424 Bia\l ystok,
Poland\\ e-mail: ep@alpha.uwb.edu.pl \\[1ex]
 Jan S\l adkowski\\ Institute of Physics, University of Silesia, \\ Uniwersytecka
4, Pl 40007 Katowice, Poland \\ e-mail: sladk@us.edu.pl }
\begin{document}
\def\meter{\mbox{$\frown\hspace{-.9em}{\lower-.4ex\hbox{$_\nearrow$}}$}}
\def\circs{\mbox{$\hspace{-.3em}\bigcirc\hspace{-.69em}{\lower-.35ex\hbox{$_{S}$}}$}}
\def\circh{\mbox{$\hspace{-.16em}\bigcirc\hspace{-.76em}{\lower-.35ex\hbox{$_{H}$}}$}}
\maketitle
\begin{center}

\begin{minipage}{5in}
\centerline{{\sc Abstract}}
\medskip

\noindent  Attention to the very physical aspects of
 information characterizes  the current research in quantum
 computation, quantum cryptography and quantum communication.  In
 most of the cases quantum description of the system provides
 advantages over the classical approach. Game theory, the study of
 decision making in conflict situation has already been extended to
 the quantum domain.  We would like to review the latest
 development in quantum game theory  that is relevant to
 information processing. We will begin by  illustrating the general
 idea of a quantum game and methods of gaining an advantage over
 ``classical opponent''. Then we  review the most important game
 theoretical aspects of quantum information processing. On grounds of the discussed material, we reason about
possible future development of quantum game theory and its impact
on information processing and  the emerging information society.
The idea of quantum artificial intelligence is explained.
\end{minipage}
\end{center}

{\it Keywords and phrases}\/: quantum games, quantum strategies,
quantum information theory, quantum computations, artificial
intelligence \\ {\it PACS Classification}\/: 02.50.Le, 03.67.Lx,
05.50.+q, 05.30.-d \\ {\it Mathematics Subject Classification}\/:
81-02, 91-02, 91A40, 81S99\\

 \vspace{1cm}


\section{Introduction}
Various sorts of computations permeate everyday life. As the
leading paradigm of computation is shifting from centralized and
static to  distributed, both in time and space, or even mobile
game theoretical methods are becoming more end more important.
Though classical computing is an extraordinary success story we
have arrived to the verge of questioning classical computational
paradigms. Since the publication of G\"{o}del theorems
\cite{nagel} and Turing and Church models of computation
\cite{Penrose} the opinion that human mind dominates any
conceivable computer prevails. But in the light of quantum
information processing \cite{NC} and scepticism concerning the
role of quantum phenomena in brain processes \cite{penrose2} we
might be doomed to dreary future of coherent states of quantum
matter dominating human mind.  A new fascinating field of research
has been started. Computational processes often take the form
agents predicting and analyzing their interactions and lead into
the domain od game theory. Quantum game theory
\cite{Mey}-\cite{inv} emerged as a valuable tool in this field
because a substantial part of front problems can be formulated in
game theoretical terms. In this paper we would like to convince
the reader that the research on quantum game theory cannot be
neglected because present technological development suggest that
sooner or later someone would take full advantage of quantum
theory and may use quantum strategies to beat us at some realistic
game. At present, it is difficult to find out if human
consciousness explores quantum phenomena although it seems to be
at least as mysterious as the quantum world. Humans have been
applying quantum technologies more or less successfully since its
discovery. Does it mean that our intelligence is being transformed
into quantum artificial intelligence (cf. quantum anthropic
principle as formulated in \cite{qantpri})? Humans have already
overcome several natural limitations with help of artificial
tools. Is quantum information processing waiting for its turn? To
exploit emerging novel nonclassical computational paradigms we
must seek for them such rigor as is possible. What that science
will look like is currently unclear, and  it is difficult to
predict which results would turn out to be fruitful and which
would have only marginal effect. The results of the research may
found applications in quantum information and cryptography, social
sciences, biology and economics.
\section{Programmable quantum systems}
The ultimate goal of quantum technology is the ability of building
quantum systems with desired (controlled) properties. We will call
them {\em programmable quantum systems} (PQS). These include
systems that can perform various computational tasks. Therefore
considerable effort has been devoted to investigating how to
efficiently control the dynamics of quantum systems and obtain and
process information on the quantum level \cite{control}. We
believe the best solution is that of building up complex behaviors
out of simple operations\footnote{Such a division is not unique
and there are models for quantum computing that have no natural
into parts at all \cite{zadoyan}. }. To this end we need
specialized systems that interact with a given quantum system to
observe and control it. Lloyd, Landahl and Slotine have described
a simple quantum devise --  a universal quantum interface that is
able to perform such tasks simply and effectively
\cite{interface}. The universal quantum interface $Q$ consists of
a single two-state quantum system (qubit) that couples to a system
$S$ whose dynamics is governed by a Hamiltonian $H$ to be
controlled or observed. The control is implemented  via a
Hamiltonian interaction of the form $A\otimes \sigma _{z}$, where
$A$ is an Hermitian operator on $S$ and $\sigma_{z}$ is the $z$
Pauli matrix. They assume that  both measurements in the
eigenvector basis of $\sigma_{z}$ on $Q$ and application of the
Hamiltonian $\gamma \sigma$, where $\gamma$ is an arbitrary real
parameter, to $Q$ can be efficiently performed. The ability to
measure with respect to the eigenvector basis combined with the
ability to perform arbitrary rotations by turning on and off
various $\gamma \sigma$ implies the ability to measure with
respect to arbitrary basis. It follows that it is possible to
apply any evolution operator of the form
$exp(-G\otimes\sigma_{x}t)$, where $G$ is an arbitrary Hermitian
operator on $S$ and the $x$ Pauli matrix acts on $Q$. If the
interface is initially prepared in the eigenstate $|+1 \rangle\ $
of $\sigma_{z}$ then the evolution together with measurement of
$Q$ in the eigenvector basis $(|-1\rangle,|+1\rangle)$ effects the
generalized "yes-no" measurement on $S$ because the system evolves
from $\rho _{S}(0)$ into either $\rho^{+}=cos(\gamma Gt)\rho
_{S}(0) cos(\gamma Gt)$ or $\rho^{-} _{S}=sin(\gamma Gt)\rho
_{S}(0)sin(\gamma Gt)$ with probabilities $P_{+}=tr\
cos^{2}(\gamma Gt)\rho _{S}(0)$ and $P_{-}=tr\ sin^{2}(\gamma
Gt)\rho _{S}(0)$, respectively. The efficiency and faithfulness of
this procedures can be analyzed analogously to the Solovay-Kitaev
theorem \cite{NC}. In general, a multiple quantum interfaces would
be necessary for efficient connection and controlling even a small
number of quantum systems. The universal quantum interface can
control a given quantum system, observe it and communicate between
various systems. It can be also used to protect the system from
disturbances \footnote{Physically realizable quantum programable
systems operate in a regime of extreme sensitivity to decoherence
and disturbances \cite{raedt}. Therefore some extra tools would be
necessary to increase reliability and stability of quantum
circuits. Quantized Parrondo's paradox \cite{Abb} and quantum Zeno
effect might turn out useful.}. Together with quantum error
correcting systems it can be used to engineer quantum subsystems
in any desired fashion \cite{interface}. Unfortunately, this does
not imply that such a system would not be extremely complex.
Therefore to implement a general purpose information processing
machine we need a possibly minimal
 set of universal quantum information processing systems that can be
connected and controlled by quantum interfaces to perform a given
task. Such universal systems do exists -- they are called
universal gates or universal primitives. But, as we would like to
focus on game-theoretical aspects, before proceeding to universal
gates we give a short introduction to quantum games.
\section{Quantization of games}
Classical games usually cannot be quantized in a unique way
because they are only asymptotical ``shadows'' of a wide spectra
of quantum models. There are two obvious modifications of
classical simulation games.
\begin{description}
\item[1 -- prequantization:] Redefine the game so that it becomes
a reversal operation on qubits representing player's strategies.
This already allows for quantum coherence of
strategies\footnote{This may result from nonclassical initial
strategies or classically forbidden measurements of the state of
the game (end of the game).}. \item[2 -- quantization:] Reduce
\label{punktdwa} the number of qubits and allow arbitrary
unitary\footnote{At least one of the performed (allowed)
operations should not be equivalent to a classical one. Otherwise
we would get a game equivalent to some variant of the prequantized
classical game.} transformation so that the basic features of the
classical game are preserved. At this stage ancillary qubits can
be introduced so that possibly all quantum subtleties can be
explored (e.g.~entanglement, measurements and the involved
reductions of states, nonlocal quantum gates etc.).
\end{description}
Basically, any quantum system that can be manipulated  by at least
one party and where the utility of the moves can be reasonably
defined, quantified and ordered may be conceived as a quantum
game. The quantum system may  be referred to as a {\em quantum
board}\/ although the term {\em universum of the game}\/ seems to
be more appropriate. We will suppose that all players know the
state of the game at the beginning and  at some crucial stages
that may depend an the game being played. This is a subtle point
because it is not always possible to identify the state of a
quantum system let alone the technical problems with actual
identification of the state (one can easily give examples of
systems that are only partially accessible to some players ). A
``realistic'' quantum game should include measuring apparatuses or
information channels that provide information on the state of the
game at crucial stages and specify the way of its termination. We
will neglect these nontrivial issues here. Therefore we can
suppose that a {\em two--player quantum game}\/
$\Gamma\negthinspace =\negthinspace({\cal
H},\rho,S_A,S_B,P_A,P_B)$ is completely specified by the
underlying Hilbert space ${\cal H}$ of the physical system, the
initial state $\rho\negthinspace\in\negthinspace {\cal S}({\cal
H})$, where ${\cal S}({\cal H})$ is the associated state space,
the sets $S_A$ and $S_B$ of permissible quantum operations of the
two players, and the { pay--off (utility) functions}\/ $P_A$ and
$P_{B\/}$, which specify the pay--off for each player. A {\em
quantum strategy}\/ $s_A\negthinspace\in\negthinspace S_A$,
$s_B\negthinspace\in\negthinspace S_B$ is a collection of
admissible quantum operations, that is the mappings of the space
of states onto itself. One usually supposes that they are
completely positive trace--preserving maps. The quantum game's
definition may also include certain additional rules, such as the
order of the implementation of the respective quantum strategies
or restriction on the admissible communication channels, methods
of stopping the game etc. We also exclude the alteration of the
pay--off during the game. The generalization for the N players
case is obvious. The real challenge is  to describe quantum games
with unlimited and changing number of players. The players should
be able to change their strategies during the course of the game
(tactics)\footnote{In the standard matrix formulation of the game
all strategies are listed and defined at the beginning. We would
like to describe more general situations, where the player can
change his mind, and, for example,  instead of buying sells some
financial asset. To this aim tactics changing strategies are
necessary.}. A possible approach is as follows. If a game allows a
great number of players in it is useful to consider it as a
two--players game: the $k$-th trader against the Rest of the World
(RW). Any concrete algorithm $\mathcal{A}$ should allow for an
effective strategy of RW (for a sufficiently large number of
players the single player strategy should not influence the form
of the RW strategy). Tactics and moves are performed by unitary
transformations on vectors in the Hilbert space (states). This
approach is suitable for description of quantum market games
\cite{PS1}. Let the real variable $q$
$$q:= \ln c - E(\ln c) $$ denotes the logarithm of the price at
which the $k$-th player can buy the asset $\mathfrak{G}$ shifted
so that its expectation value in the state $|\psi\rangle_{k}$
vanishes. The expectation value of $x$ is denoted by $E(x)$. The
variable $p$
$$p:= E(\ln c) - \ln c  $$ describes the situation of a player who
is supplying the asset $\mathfrak{G}$ according to his strategy
$|\psi\rangle_k$. Supplying $\mathfrak{G}$ can be regarded as
demanding $\$$ at the price $c^{-1}$ in the $1\mathfrak{G}$ units
and both definitions are equivalent. Note that we have defined $q$
and $p$ so that they do not depend on possible choices of the
units for $\mathfrak{G}$ and $\$ $. For simplicity we will use
such units that $E(\ln c) =0$. The strategies $|\psi \rangle_{k}$
belong to  Hilbert spaces $H_{k}$. The state of the game
$|\Psi\rangle_{in}:=\sum_k|\psi \rangle_k$ is a vector in the
direct sum of Hilbert spaces of all players. We
 define canonically conjugate Hermitian operators of demand
$\mathcal{Q}_k$ and supply $\mathcal{P}_k$ for each Hilbert space
$H_{k}$ analogously to their physical position and momentum
counterparts. This can be justified in the following way. Let
$\text{e}^{-p}$ be a definite price, where $p$ is a proper value
of the operator $\mathcal{P}_k$. Therefore, if one have already
declared the will of selling exactly at the price $\text{e}^{-p}$
(the strategy given by the proper state $|p\rangle_{k}$) then it
is pointless to put forward any opposite offer for the same
transaction. The capital flows resulting from an ensemble of
simultaneous transactions correspond to the physical process of
measurement. A transaction consists in a transition from the state
of traders strategies $|\Psi\rangle_{in}$ to the one describing
the capital flow state $|\Psi\rangle_{out}:=\mathcal{T}_\sigma
|\Psi\rangle_ {in}$, where
$\mathcal{T}_{\sigma}:=\sum_{k_d}|q\rangle_{k_d}\phantom{}_{k_d}
   \negthinspace\langle q|+
 \sum_{k_s}|p\rangle_{k_s}\phantom{}_{k_s}
   \negthinspace\langle p|$  is the projective operator defined by
the division $\sigma $ of the set of traders $\{ k\}$ into two
separate subsets $\{k\}=\{k_d\}\cup\{k_s\}$, the ones buying at
the price $\text{e}^{q_{k_d}}$ and the ones selling at the price
$\text{e}^{-p_{k_s}}$ in the round of the transaction in question.
The role of the algorithm $\mathcal{A}$ is to determine the
division $\sigma$ of the market, the set of price parameters $\{
q_{k_{d}}, p_{k_{s}}\}$ and the values of capital flows. The later
are settled by the distribution $$\int_{-\infty}^{\ln c}
\frac{{|\langle q|\psi\rangle_k|}^2}{\phantom{}_k
\negthinspace\langle\psi|\psi\rangle_k}dq $$ which is interpreted
as the probability that the trader $| \psi \rangle _{k}$ is
willing to buy  the asset $\mathfrak{G}$ at the transaction price
$c$ or lower. In an analogous way the distribution $$
\int_{-\infty}^{\ln \frac{1}{c}} \frac{{|\langle
p|\psi\rangle_k|}^2}{\phantom{}_k
\negthinspace\langle\psi|\psi\rangle_k}dp  $$ gives the
probability of selling $\mathfrak{G}$ by the trader $| \psi
\rangle _{k}$ at the price $c$ or greater. These probabilities are
in fact conditional because they describe the situation after the
division $ \sigma $ is completed. If one considers the RW strategy
it make sense to declare its simultaneous demand and supply states
because for one player RW is a buyer and for another it is a
seller. To describe such situation it is convenient to use the
Wigner formalism\footnote{Actually, this approach consists in
allowing pseudo--probabilities into consideration. From the
physical point of view this is questionable but for our aims its
useful, cf\mbox{.} the discussion of the Giffen paradox
\cite{Sla}.}. The pseudo--probability $W(p,q)dpdq$ on the phase
space $\{(p,q)\}$ known as the Wigner function is given by
\begin{equation*}
W(p,q):= h^{-1}_E\int_{-\infty}^{\infty}\text{e}^{i\hslash_E^{-1}p
x} \;\frac{\langle
q+\frac{x}{2}|\psi\rangle\langle\psi|q-\frac{x}{2}\rangle}
{\langle\psi|\psi\rangle}\; dx .
\end{equation*}
In general, this measure is not positive definite. In the general
case the pseudo--probability density of RW is a countable linear
combination of Wigner functions, $\rho(p,q)=\sum_n w_n W_n (p,q)$,
 $w_n\geq 0$, $\sum_n w_n =1$.\\

\begin{figure}[h]
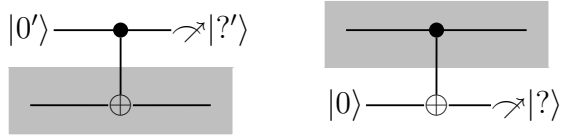

\begin{center}
\phantom{a}\vspace{5ex} \psset{linewidth=.7pt}
\rput(-3.3,1){\rnode{A}{$|0'\rangle$\hspace{1pt}}}
\cnode*(-2.1,1){.1}{B}
\rput(-0.9,1){\rnode{C}{\meter$|?'\rangle$}}
\pscustom[linecolor=white,fillstyle=solid,fillcolor=lightgray]{%
\psline(3.6,.5)(3.6,.5) \psline(3.6,1.4)(.6,1.4)
\psline(.6,1.4)(.6,.5) } \rput(0.9,1){\rnode{D}{}}
\cnode*(2.1,1){.1}{E} \rput(3.3,1){\rnode{F}{}}
\rput(-3.3,0){\rnode{G}{}}
\pscustom[linecolor=white,fillstyle=solid,fillcolor=lightgray]{%
\psline(-3.6,.5)(-3.6,.5) \psline(-3.6,-.4)(-.6,-.4)
\psline(-.6,-.4)(-.6,.5) } \rput(-2.1,0){\rnode{H}{$\oplus$}}
\rput(-0.9,0){\rnode{I}{}}
\rput(0.9,0){\rnode{J}{$|0\rangle$\hspace{1pt}}}
\rput(2.1,0){\rnode{K}{$\oplus$}}
\rput(3.3,0){\rnode{L}{\meter$|?\rangle$}}
\ncline[nodesep=0pt]{-}{A}{C} \ncline[nodesep=0pt]{-}{D}{F}
\ncline[nodesep=0pt]{-}{G}{H} \ncline[nodesep=0pt]{-}{H}{I}
\ncline[nodesep=0pt]{-}{J}{K} \ncline[nodesep=0pt]{-}{K}{L}
\ncline[nodesep=0pt]{-}{B}{H} \ncline[nodesep=0pt]{-}{E}{K}
\end{center}
\caption{The alliance as a mean of determining others' strategies.
The sign ``\meter\hspace{.06em}'' at the right ends of lines
representing qubits symbolizes measurement.} \label{sulokitek}
\end{figure}

\begin{figure}[h]
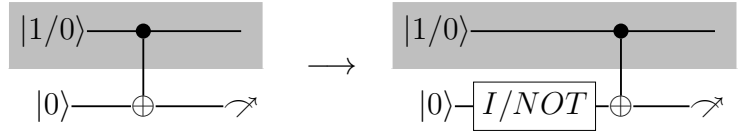

\phantom{.}\vspace{5ex}
\begin{center}
\pscustom[linecolor=white,fillstyle=solid,fillcolor=lightgray]{%
\psline(-4.9,.5)(-4.9,1.4) \psline(-4.9,1.4)(-1.5,1.4)
\psline(-1.5,.5)(-1.5,.5) }
\pscustom[linecolor=white,fillstyle=solid,fillcolor=lightgray]{%
\psline(0.2,.5)(0.2,.5) \psline(0.2,1.4)(4.8,1.4)
\psline(4.8,1.4)(4.8,.5) } \rput(-4.3,1){\rnode{A}{$|1/0\rangle$}}
\cnode*(-3.1,1){.1}{B} \rput(-1.8,1){\rnode{C}{}}
\rput(-4.3,0){\rnode{D}{$|0\rangle$}}
\rput(-3.1,0){\rnode{E}{$\oplus$}}
\rput(-1.8,0){\rnode{F}{\meter}}
\rput(-.6,.5){\rnode{G}{$\longrightarrow$}}
\rput(.8,1){\rnode{H}{$|1/0\rangle$}} \cnode*(3.25,1){.1}{I}
\rput(4.5,1){\rnode{J}{}} \rput(.8,0){\rnode{K}{$|0\rangle$}}
\rput(2.1,0){\rnode{L}{\psframebox[linewidth=.3pt]{$I/NOT$}}}
\rput(3.25,0){\rnode{M}{$\oplus$}} \rput(4.4,0){\rnode{O}{\meter}}
\psset{linewidth=.7pt} \ncline[nodesep=0pt]{-}{A}{C}
\ncline[nodesep=0pt]{-}{D}{E} \ncline[nodesep=0pt]{-}{E}{F}
\ncline[nodesep=0pt]{-}{B}{E}
\ncline[nodesep=0pt]{-}{H}{J} \ncline[nodesep=0pt]{-}{K}{L}
\ncline[nodesep=0pt]{-}{L}{M} \ncline[nodesep=0pt]{-}{M}{O}
\ncline[nodesep=0pt]{-}{I}{M}
\end{center}
\caption{Neutralization  of a quantum measuring system by a switch
$I/NOT$ applied ($I/NOT\rightarrow NOT$), when
$|1/0\rangle\negthinspace=\negthinspace|1\rangle$ (see the text).}
\label{grasweta}
\end{figure}

 One of the most appealing features of quantum games is
the possibility that  strategies can influence each other and form
collective strategies. Elsewhere \cite{komp}, we have defined the
alliance as  the gate CNOT ($\mathcal{C}$) regardless of its
standard name {\em controlled-NOT}\/  because it can be used to
form collective strategies as follows. Most of two-qubit quantum
gates are universal in the sense that any other gate can be
composed of a universal one \cite{DBE}-\cite{Lom}. Therefore it is
sufficient to describe a collective tactic of $N$ players as a
sequence of various operations $\mathcal{U}_{z,\alpha}$ belonging
to $SU(2)$ performed on one-dimensional subspaces of players'
strategies and, possibly, alliances $\mathcal{C}$ among them (any
element of $SU(2^N)$ can be given such a form \cite{BBC}).
Alliances are, up to equivalence, the only ways of forming
collective games. An alliance has the explicit form
$CNOT:=|0\rangle\langle0|\otimes I+ |1\rangle\langle1|\otimes
NOT$, where the tactic $NOT$ is represented in the qubit basis
$(|0\rangle,|1\rangle)$ by the matrix $\begin{pmatrix}0 &
\text{i}\\\text{i}&0\end{pmatrix}\negthinspace\in\negthinspace
SU(2)$. An alliance allows the player to determine the  state of
another player by entering into an alliance and measuring her
resulting strategy. This process is shortly described as
\begin{equation*}
\mathcal{C}\,|0'\rangle|m'\rangle=|m'\rangle|m'\rangle,\,\,\,\,
\mathcal{C}\,|m\rangle|0\rangle=|m\rangle|m\rangle,
\end{equation*}
where $m\negthinspace=\negthinspace0,\text{1}$. The corresponding
diagrams are shown in Fig\mbox{.} \ref{sulokitek}. The left
diagram presents measurement of the observable $\mathcal{X}'$ and
the right one  measurement of $\mathcal{X}$.
Any measurement would demolish possible entanglement of
strategies.  Therefore entangled quantum strategies can exist only
if the players in question  are ignorant of the details of their
strategies. To illustrate the problem we analyze three simple
games involving alliances. They can be used as partial solutions
in more complicated situations. To taste power of the formalism
let us investigate  the  Newcomb's paradox \cite{PSN}. Any circuit
is more or less vulnerable to random errors. Consider the simple
quantum circuit presented in Fig\mbox{.} \ref{grasweta}. The gate
$I/NOT$ is defined as a randomly chosen gate from the set
$\{I,NOT\}$) and is used to switching-off the circuit in a random
way. It can be generalized to have some additional control qubits.
In a game-theoretical context such circuits can be used to
neutralization of disturbances caused by measuring strategies,
c.f. \cite{NC}.
For example, it can be applied to solve the famous Newcomb's free
will paradox. The problem, originally formulated by William
Newcomb in the 1960, was described by Martin Gardner in the
following way \cite{Gar}.  An alien Omega being a representative
of alien civilization (player 2) offers a human (player 1) a
choice between two boxes. The player 1 can take the content of
both boxes or only the content of the second one. The first one is
transparent and contains \$1000. Omega declares to have put into
the second box that is opaque \$1000000 (strategy
$|\mathsf{1}\rangle_2$) but only if Omega foresaw that the player
1 decided to take only the content of that box
($|\mathsf{1}\rangle_1$). A male player 1 thinks: {\em If Omega
knows what I am going to do then I have the choice between \$1000
and \$1000000. Therefore I take the \$1000000 }(strategy
$|\mathsf{1}\rangle_1$). A female player 1 thinks: {\em Its
obvious that I want to take the only the content of the second box
therefore Omega foresaw it and put the \$1000000 into the box. So
the one million dollar {\bf is} in the second box. Why should I
not take more -- I take the content of both boxes} (strategy
$|\mathsf{0}\rangle_1$). The question is whose strategy, male's or
female's, is better? In he measuring system presented in
Fig\mbox{.} \ref{grasweta} the initial value $|0\rangle$ of the
lower qubit corresponds to the male strategy and the values
$|1\rangle$ and $|0\rangle$ of the upper qubit correspond to male
and female tactics, respectively. The outcome $|0\rangle$ of a
measurement performed on the lower qubit indicates the opening of
both boxes with contents prepared by Omega before the alliance
$CNOT$ was formed. If Omega  installed in the circuit a breaker of
the form $I/NOT$ (before or after the alliance $CNOT$ he would use
it when (and only then) the human adopted the female tactics. But
this would mean that Omega is cheating (the breaker is installed
after the alliance) or is able to foretell the future (the breaker
is installed before the alliance). In the quantum setting the
situation is different.
\begin{figure}[h]
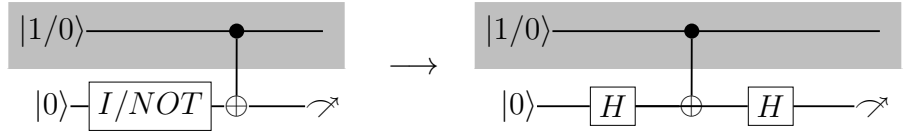

\phantom{.}\vspace{5ex}
\begin{center}
\pscustom[linecolor=white,fillstyle=solid,fillcolor=lightgray]{%
\psline(-6,.5)(-6,1.4) \psline(-6,1.4)(-1.5,1.4)
\psline(-1.5,.5)(-1.5,.5) }
\pscustom[linecolor=white,fillstyle=solid,fillcolor=lightgray]{%
\psline(.2,.5)(.2,.5) \psline(.2,1.4)(5.9,1.4)
\psline(5.9,1.4)(5.9,.5) }
\rput(-5.4,1){\rnode{A}{$|1/0\rangle$}} \cnode*(-2.95,1){.1}{B}
\rput(-1.8,1){\rnode{C}{}} \rput(-5.4,0){\rnode{D}{$|0\rangle$}}
\rput(-4.1,0){\rnode{E}{\psframebox[linewidth=.3pt]{$I/NOT$}}}
\rput(-2.95,0){\rnode{F}{$\oplus$}}
\rput(-1.8,0){\rnode{G}{\meter}}
\rput(-.6,.5){\rnode{H}{$\longrightarrow$}}
\rput(.8,1){\rnode{I}{$|1/0\rangle$}} \cnode*(3.1,1){.1}{J}
\rput(5.6,1){\rnode{K}{}} \rput(.8,0){\rnode{L}{$|0\rangle$}}
\rput(2.05,0){\rnode{M}{\psframebox[linewidth=.3pt]{$H$}}}
\rput(3.1,0){\rnode{N}{$\oplus$}}
\rput(4.15,0){\rnode{O}{\psframebox[linewidth=.3pt]{$H$}}}
\rput(5.5,0){\rnode{P}{\meter}}
\psset{linewidth=.7pt} \ncline[nodesep=0pt]{-}{A}{C}
\ncline[nodesep=0pt]{-}{D}{E} \ncline[nodesep=0pt]{-}{E}{F}
\ncline[nodesep=0pt]{-}{F}{G} \ncline[nodesep=0pt]{-}{B}{F}
\ncline[nodesep=0pt]{-}{I}{K} \ncline[nodesep=0pt]{-}{L}{M}
\ncline[nodesep=0pt]{-}{M}{N} \ncline[nodesep=0pt]{-}{M}{O}
\ncline[nodesep=0pt]{-}{O}{P} \ncline[nodesep=0pt]{-}{J}{N}
\end{center}
\caption{Solution to the Newcomb's paradox: quantum device that
neutralizes measurement. In the quantization process the gate
$I/NOT$ is replaced by a qutrojan (see the text) that acts
independently of the value of the qubit $|1/0\rangle$ and is
composed of  two Hadamard gates $H$.} \label{graswetcc}
\end{figure}
The quantization of the problem is presented in
Fig.~\ref{graswetcc}. It consists in replacing of the
circuit-breaker $I/NOT$ by a pair of Hadamard gates
$H:=\frac{\text{i}}{\sqrt{2}}
\begin{pmatrix}1&\phantom{-}1\\
1&-1\end{pmatrix}\negthinspace\in\negthinspace SU(2)$. Due to
their jamming effect on the human's tactics, we can call them a
quantum Trojan horse (qutrojan)\footnote{Problems connected with
the definition of trojan are discussed in \cite{TAC}.}. We can
hardy use the term trojan with respect to the circuit-breaker
$I/NOT$ because of its paradoxical correlation with human tactics.
Note that
$H\cdot NOT \cdot H= \begin{pmatrix}-\text{i}&0\\
\phantom{-}0&\text{i}\end{pmatrix}$, hence any attempt at
measuring squared absolute values of coordinates of the human
strategy qubit will not detect any effectiveness of the female
tactics.
\section{Universal primitives of quantum games}
There are two main approaches to the universality problem of
quantum computation. First approach consist in approximation of a
unitary transformation $U=\exp(-iHt)$ in a way analogous to
infinitesimal operators in Lie group theory \cite{DBE,vin1,Lloyd}.
In the alternative approach one tries to represent the matrix of a
given unitary transformation  as a product of one- and two-qubit
gates from a possibly minimal set - set of {\em universal
primitives}. A related approach follows the methods used in
teleportation -- the dominant role play measurements. Raussendorf
and Briegel \cite{raus, Nielsen-c} proposed the so called
cluster-state model that forms a powerful tool in quantum
complexity theory \cite{raus2}. A more practical approach to the
measurement-based model of quantum computation was proposed by
Nielsen \cite{Nielsen,NC}.  This model of quantum computation has
been further developed by Leung \cite{Leung1} and Leung and
Aliferis \cite{Leung2}  who exhibited a universal family of
universal primitives composed of $4$ two-qubit measurements ($4$
auxiliary qubits are necessary). The most important result in
measurement-based models of computation was obtained by Perdix
\cite{Per} who has introduced a model of measurement-based quantum
computation
 that makes use of what he calls the {\em state
transfer}. In this model of quantum computation  to simulate any
$n$-qubit unitary transformation one auxiliary qubit is required
-- a universal family of observables could be formed by $3$
one-qubit measurements and only one two-qubit measurement.
Perdrix's \cite{Per} approach improves previous results by
reducing both the number of auxiliary qubits and the number of
two-qubit measurements required for quantum universality. From the
theoretical point of view\footnote{We neglect here  the problem of
optimal convergence. Other classes of universal primitives are
usually introduced to analyze  this nontrivial issue. For example,
a set of quantum gates is said to be {\em computationally
universal} if it can be used to simulate to within $\varepsilon$
error any quantum circuit which uses $n$ qubits and $t$ gates from
a (strictly) universal set with only polynomial overheads in
$(n,t,\frac{1}{\varepsilon})$. A non-minimal set might be more
effective in simulations \cite{motto, Nielsen-g}} the minimal
amounts of necessary resources are reached in this approach. It
follows that quantum games with a ``fixed quantum board'' that is
implemented as a fixed quantum circuit have the same universal
properties as the circuit used for their implementation. In an
open dynamical setting, such as in quantum market games the
situation is slightly more complicated but a concise dictionary of
universal primitives can be given \cite{pps1} if one follows the
way paved by Perdix \cite{Per, jorrand1}. In this case
measurements also form sufficiently powerful and effective tools
in manipulation of quantum tactics. A measurement of tactics
consists in determination of the strategy or, more precise,
finding out which of its fixed points we have to deal with.  If
the tactics being measured changes the corresponding strategy,
then the non-demolition measurement reduces the strategy to one of
its fixed points and the respective transition amplitudes are
given by coordinates of the strategy in the fixed point basis.  A
dominant role of measurements in implementations of this type
suggests that quantum games may be free from psychological
factors, such as phobia, intention, irrationality and so forth.
Following Perdrix \cite{Per,pps1}, we can see that measurements of
the tactics $X$, $G$ and $X\otimes X'$, where
\[
X:=\sigma_x\,,\,\,X':=\sigma_z=HXH\,,\,\,
 G:=\tfrac{1}{\sqrt{2}}(X'+X'')\,.
\]
 suffice to implement quantum market
games. Graphically, these measurements will be represented as
\footnote{We follow the {\em Qcircuit.tex}\/ convention
\cite{eastin}. Thus  rounded off shape is used to distinguish
measuring gates.}.
\begin{equation} \Qcircuit @C=1em @R=.7em {
 &\bgate{X}&\qw
} \,\,,\,\, \Qcircuit @C=1em @R=.7em {
 &\bgate{G}&\qw
} \,\,,\,\, \lower-.9em\hbox{\Qcircuit @C=1em @R=.7em{
&\bmultigate{1}{X\negthinspace\otimes\negthinspace X'}&\qw\\
&\ghost{X\negthinspace\otimes\negthinspace X'}&\qw }}\,\,.
\label{uniwersalne}
\end{equation}
Measurement of the tactics $X\negthinspace\otimes\negthinspace X'$
provides us with information whether the two strategies  agree or
disagree on the price but reveals no information on the level of
the price in question. To get information about the prices we have
to measure $X\otimes I$ and $I\otimes X'$, respectively. Note that
the measurement of $X'$ can be accomplished implicitly  by
measurement of $X$ and subsequently $X\otimes X'$. Graphically
this is represented as \cite{Per, jorrand1}:
\[
\Qcircuit @C=1em @R=.7em{
{\scriptstyle(}&\bgate{X}&\bmultigate{1}{{X\negthinspace\otimes\negthinspace
X'}}&
\qw {\scriptstyle)}\\
&\qw&\ghost{X\negthinspace\otimes\negthinspace X'}&\qw }\,\,
\lower.7em\hbox{\,\,\,\,$\Longrightarrow$\,\,\,} \lower.4em\hbox{
\,\,\Qcircuit @C=1em @R=.7em { &\bgate{X'}&\qw }\,\, ,}
\]
where the parentheses are used to denote auxiliary qubits. The
following sequence of measurements shows that a strategy encoded
in one qubit can be transferred to another qubit (from the upper
one to the lower one in the figure bellow) and changed with the
 tactics $\sigma H$, where
 $\sigma$ is one of the
Pauli matrices (including the identity matrix):
\begin{equation}
\Qcircuit @C=1em @R=.7em{
&\qw&\bmultigate{1}{{X\negthinspace\otimes\negthinspace X'}}&
\bgate{X'}&\qw {\scriptstyle)}\\
{\scriptstyle(}&\bgate{X}&\ghost{X\negthinspace\otimes\negthinspace
X'}&\qw&\qw }\,\, \lower.7em\hbox{\,\,\,\,$\Longrightarrow$\,\,\,}
\lower.4em\hbox{ \,\,\Qcircuit @C=1em @R=.7em { &\gate{\sigma
H}&\qw }\,\,.} \label{hgate}
\end{equation}

Thus the strategy encoded in the upper state  is transferred from
the lower qubit and changed with the tactics $\sigma H$. It is
obvious that the same tactics is adopted if we switch the supply
measurements with the demand ones
($X\negthinspace\leftrightarrow\negthinspace X'$). Simple
calculation shows that the composite tactics $H\sigma H$ and
$\sigma _{i}\sigma _{k}$ reduce to some Pauli (matrix) tactics.
Therefore an even sequence of tactics $(\ref{hgate})$ can be
perceived as a Markov process over vertices of the graph 
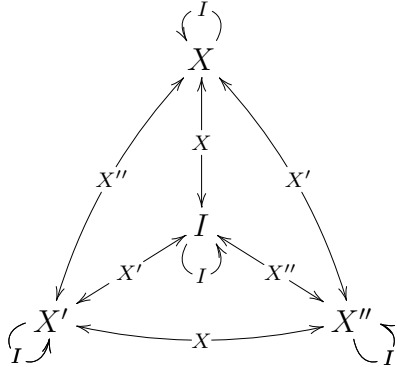
\[ \xy
(0,1)*+{I}="a",(0,23.1)*+{X}="b",(-20,-11.55)*+{X'}="c",(20,-11.55)*+{X''}="d"
\ar@(ld,rd)|{\,I\,}"a";"a" \ar@(ur,ul)|{\;I\,}"b";"b"
\ar@(l,d)|{\;I\,} "c";"c" \ar@(d,r)|{\vphantom{^I}\;I\,}"d";"d"
\ar@{<->}@/^1.5ex/|{\,\,\vphantom{\sum^I}X'}"b";"d"
\ar@{<->}@/_1.5ex/|{\vphantom{\sum^I}X''\negthinspace\negthinspace}"b";"c"
\ar@{<->}|{\vphantom{\sum^I}X}"b";"a" \ar@(l,d)|{\;I\,}"c";"c"
\ar@{<->}|{\,\,\vphantom{\sum^I}X'}"c";"a"
\ar@(d,r)|{\vphantom{^I}\;I\,}"d";"d"
\ar@{<->}|{\vphantom{\sum^I}X''\negthinspace\negthinspace}"d";"a"
\ar@{<->}@/^1.5ex/|{\vphantom{\sum^I}X}"d";"c"
\endxy
\]
It follows that  any Pauli tactics can be implemented  as an even
number of tactics-measurements $(\ref{hgate})$ by identifying it
with some final vertex of random walk on this graph. Although the
probability of drawing out the final vertex at the first step is
$\tfrac{1}{4}$, the probability of staying in the ``labirynth''
decreases to zero exponentially with time. Having a method of
implementation of Pauli tactics allows us to modify the tactics
$(\ref{hgate})$ so that to implement the tactics $H$ --- the
fundamental operation of switching the supply representation with
the demand representation. It can be also used to measure
compliance with tactics representing the same side of the market
(direct measurement is not possible because  the agents cannot
make the deal):
\[
\Qcircuit @C=1em @R=.7em{
&\bmultigate{1}{X\negthinspace\otimes\negthinspace X}&\qw\\
&\ghost{X\negthinspace\otimes\negthinspace X}&\qw }
\lower.6em\hbox{\,\,\,$:=$\,\,\,} \Qcircuit @C=1em @R=.7em{
&\qw&\bmultigate{1}{X\negthinspace\otimes\negthinspace X'}&\qw&\qw\\
&\gate{H}&\ghost{X\negthinspace\otimes\negthinspace
X'}&\gate{H}&\qw }\,\,\lower.6em\hbox{\,,}
\]
\[
\Qcircuit @C=1em @R=.7em{
&\bmultigate{1}{X'\negthinspace\otimes\negthinspace X'}&\qw\\
&\ghost{X'\negthinspace\otimes\negthinspace X'}&\qw }
\lower.6em\hbox{\,\,\,$:=$\,\,\,} \Qcircuit @C=1em @R=.7em{
&\gate{H}&\bmultigate{1}{X\negthinspace\otimes\negthinspace X'}&\gate{H}&\qw\\
&\qw&\ghost{X\negthinspace\otimes\negthinspace X'}&\qw&\qw
}\lower1em\hbox{\,\,\,.}
\]
In addition, this would allow for interpretation via measurement
of random Pauli tactics $\sigma$ because due to the involutiveness
of $H$ the gate $(\ref{hgate})$ can be transformed to
\begin{equation}
\Qcircuit @C=1em @R=.7em{
&\gate{H}&\bmultigate{1}{{X\negthinspace\otimes\negthinspace X'}}&
\bgate{X'}&\qw {\scriptstyle)}\\
{\scriptstyle(}&\bgate{X}&\ghost{X\negthinspace\otimes\negthinspace
X'}&\qw&\qw }\lower.8em\hbox{\,\,\,\,$=$\,\,\,\,} \Qcircuit @C=1em
@R=.7em{ &\qw&\bmultigate{1}{{X'\negthinspace\otimes\negthinspace
X'}}&
\bgate{X}&\qw {\scriptstyle)}\\
{\scriptstyle(}&\bgate{X}&\ghost{X'\negthinspace\otimes\negthinspace
X'}&\qw&\qw } \lower.8em\hbox{\,\,\,\,$=$} \label{sigmagate}
\end{equation}
\[
\Qcircuit @C=1em @R=.7em{
&\qw&\bmultigate{1}{{X\negthinspace\otimes\negthinspace X}}&
\bgate{X'}&\qw {\scriptstyle)}\\
{\scriptstyle(}&\bgate{X'}&\ghost{X\negthinspace\otimes\negthinspace
X}&\qw&\qw } \,\, \lower.7em\hbox{\,\,\,\,$\Longrightarrow$\,\,\,}
\lower.4em\hbox{ \,\,\Qcircuit @C=1em @R=.7em { &\gate{\sigma}&\qw
}\,\,.}
\]
The gate $(\ref{sigmagate})$ can be used to implement the
phase-shift tactics:
\[T:=\begin{pmatrix}
  1& 0 \\
  0& \tfrac{1+\text{i}}{\sqrt{2}}
\end{pmatrix}\,.
\]
$T$ commutes with  $X'$, hence:
\[
\lower.4em\hbox{ \,\,\Qcircuit @C=1em @R=.7em { &\gate{\sigma
T}&\qw }} \lower.7em\hbox{\,\,\,\,$\Longleftarrow$\,\,\,}
\Qcircuit @C=1em @R=.7em{
&\qw&\bmultigate{1}{{X'\negthinspace\otimes\negthinspace X'}}&
\bgate{T^{-1}X\,T}&\qw {\scriptstyle)}\\
{\scriptstyle(}&\bgate{X}&\ghost{X'\negthinspace\otimes\negthinspace
X'}&\qw&\qw } \lower1em\hbox{\,\,\,.}
\]
Elementary  calculation demonstrates that
$T^{-1}XT\negthinspace=\negthinspace\tfrac{X-X''}{\sqrt{2}}$ and
$H\tfrac{X-X''}{\sqrt{2}}H\negthinspace=\negthinspace G$,
therefore
\[\Qcircuit @C=1em @R=.7em{
&\gate{H}&\bmultigate{1}{{X\negthinspace\otimes\negthinspace X'}}&
\bgate{G}&\qw {\scriptstyle)}\\
{\scriptstyle(}&\bgate{X}&\ghost{X\negthinspace\otimes\negthinspace
X'}&\qw&\qw } \,\,
\lower.7em\hbox{\,\,\,\,$\Longrightarrow$\,\,\,} \lower.4em\hbox{
\,\,\Qcircuit @C=1em @R=.7em { &\gate{\sigma T}&\qw }}
\lower1em\hbox{\,\,\,.}
\]
We have seen earlier that it is possible to remove the superfluous
Pauli operators, cf. $(\ref{sigmagate})$. To end the proof of
universality of the set of gates $(\ref{uniwersalne})$ we have to
show how to implement the alliance $Cnot$ (note that
$\{H,T,Cnot\}$ a set of universal gates \cite{bar}). This gate can
be implemented as the circuit \cite{jorrand1} (as before, the gate
is constructed up to a Pauli tactics):
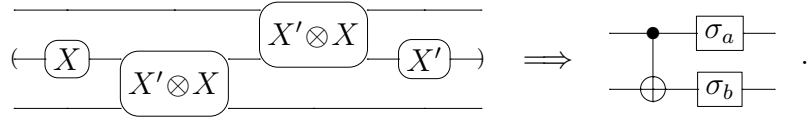
\[
\Qcircuit @C=1em @R=.7em{
&\qw&\qw&\bmultigate{1}{X'\negthinspace\otimes\negthinspace X}&\qw&\qw\\
{\scriptstyle(}&\bgate{X}&\bmultigate{1}{X'\negthinspace\otimes\negthinspace
X}&
\ghost{{X'\negthinspace\otimes\negthinspace X}}&\bgate{X'}&\qw{\scriptstyle)} \\
&\qw&\ghost{X'\negthinspace\otimes\negthinspace X}&\qw&\qw&\qw
}\,\, \lower1.6em\hbox{\,\,\,\,\,\,$\Longrightarrow$\,\,\,}
\lower.5em\hbox{ \,\,\Qcircuit @C=1em @R=.7em {
&\ctrl{1}&\gate{\sigma_a}&\qw\\
&\targ&\gate{\sigma_b}&\qw }\,\,\lower1em\hbox{\,\,\,.}}
\]
Actually, simple calculation \cite{Per} prove that the
universality property has any set of primitive that contains the
$controlled\ H$ gate and measurements $X^k,
X^p\negthinspace\otimes\negthinspace X^q,
X^r\negthinspace\otimes\negthinspace X^s$, where
$p\negthinspace\neq\negthinspace q$,
$r\negthinspace\neq\negthinspace s$ i
$p\negthinspace\neq\negthinspace r$. It follows, that to implement
a quantum market\footnote{Actually any finite-dimensional quantum
system can be implemented in that way \cite{jorrand1}.} it
suffices to have, beside possibility of measuring strategy-qubits
and control of the supply-demand context, a  direct method
measuring entanglement of a pair of qubits in conjugated bases.
The universal quantum interfaces can be used to connect and
control an a priori arbitrary number of information processing
units\footnote{The idea that a single quantum interface that can
dynamically moved between system is attractive but is probably
hard to put into effect.}. Some interesting technical details can
be found in Ref. \cite{vlasov1}. It is natural to wonder how small
such a primitive processing unit could be.  A single atom or a
molecule are examples of a possible simple quantum computing units
\cite{robin} but the feasibility of framing them into an
all-purpose quantum computer is currently out of reach.
\section{Quantum gambling}
Sophisticated technologies that are not yet available are not
necessary to implement quantum games \cite{jangfeng}. Simulation
of quantum games can be performed in an analogous way to precision
physical measurements during which classical apparatuses are used
to explore  quantum phenomena. We envisage  that quantum lotteries
will soon emerge and will challenge the present day lottery market
based randomized events or pseudo-randomness. In games of chance
the player is betting in advance on the outcomes of several
incompatible measurements. Quantum phenomena offer true random
event and commercial random event generators should appear on the
market at moderate prices \cite{mapa}. At the present stage of our
technological development it already is feasible to open {\em
quantum casinos}, where gambling at quantum games would be
possible. Of course, such an enterprise would be costly but if you
recall the amount of money spent on gambling, lotteries and
advertising various products it seems to us that it is a worthy
cause. Goldenberg, Vaidman and Wiesner described the following
game based on the coin tossing protocol \cite{Gol}. Alice has two
boxes, $A$ and $B$, which can store a particle. The quantum states
of the particle in the boxes are denoted by $|a\rangle$ and
$|b\rangle$, respectively. Alice prepares the particle in some
state and sends box $B$ to Bob. Bob wins in one of the two cases:
\begin{enumerate}
\item If he finds the particle in box $B$, then Alice pays him $1$
monetary unit (after checking that box $A$ is empty). \item If he
asks Alice to send him box $A$ for verification and he finds that
she initially prepared a state  different from $ |\psi_{0}\rangle
= 1/\sqrt{2} \: (|a\rangle + |b\rangle) , $ then Alice pays him
$R$ monetary units.
\end{enumerate}
In any other case Alice wins, and Bob pays her $1$ monetary unit.
They have analyzed the security of the scheme, possible methods of
cheating and calculated the average gain of each party as a result
of her/his specific strategy. The analysis shows that the protocol
allows two remote parties to play a gambling game, such that in a
certain limit it becomes a fair game. No unconditionally secure
classical method is known to accomplish this task. This game was
implemented by Yong--Sheng Zhang et al, \cite{Zha}. Other
proposals based on properties of non--orthogonal states have been
put forward by Hwang, Ahn, and Hwang \cite{Hwa1} and  Hwang and
Matsumoto \cite{Hwa2}. Witte has proposed a quantum version of the
Heads or Tails game \cite{Wit}. Piotrowski and S\l adkowski have
suggested that although  sophisticated technologies to put  a
quantum market in motion  are not yet available, simulation of
quantum markets and auctions can be performed in an analogous way
to precision physical measurements during which classical
apparatuses are used to explore  quantum phenomena. People seeking
after excitement would certainly not miss the opportunity to
perfect their skills at  ``using quantum strategies''. To this end
an automatic quantum game  will be sufficient and such a device
can be built up  due to the recent advances in technology
\cite{komp, mapa}. Segre has published an interesting detailed
analysis of quantum casinos and a Mathematica packages for
simulating quantum gambling \cite{Seg}. He has introduced  a
quantum analogue of the Law of Excluded Gambling Strategies of
Classical Decision Theory. The necessity of keeping into account
entanglement  requires to adopt the general algebraic language of
Quantum Probability Theory. There is  a deep link between the
theory of winning quantum gambling strategies and the central
notion of Quantum Algorithmic Information Theory -- quantum
algorithmic randomness. Quantum gambling besides its commercial
significance is closely related to quantum logic, decision theory
and can be used for defining a Bayesian theory of quantum
probability \cite{Pit} -- interesting fields of research with
various possible commercial applications.
\section{Quantum combinatorial games and quantum automata}
To our knowledge,  algorithmic combinatorial games, except for
cellular automata,  have been completely ignored by quantum
physicists. This is astonishing because at least some of the
important intractable problems might be attacked and solved on a
quantum computer.  Consider some problem $X$. Let us define the
game $kXcl$: you win if and only if you solve the problem (perform
the task) $X$ given access to only $k$ bits of information. The
quantum counterpart reads: solve the problem $X$ on a quantum
computer or other quantum device given access to only $k$ bits of
information. Let us call the game $kXcl$ or $kXq$ interesting if
the corresponding limited information--tasks are feasible. Let
$OckhamXcl$ ($OckhamXq$) denotes the minimal $k$ interesting game
in the class $kXcl$ ($kXq$).  There are a lot of intriguing
questions that can be ask, for example for which $X$ the
meta--game $Ockham(OckhamXq)cl$ can be solved or when, if at all,
the meta--problem $Ockham(OckhamXq)q$ is well defined problem.
Quantum automata (quantum state machines) \cite{albert}-\cite{M-C}
play in quantum information theory  role analogous to that of
finite automata in Turing-Church model of computation. Quantum
automaton can be defined as a quadruple $A=(S, s_{0}, \alpha ,U)$,
where $S$ is the set of allowed internal states, $s_{0} \in S$ the
initial state vector $\alpha$ the input alphabet and $U$ a unitary
transition matrix for each symbol $a\in \alpha $ \cite{automaty}.
Primary interest in quantum automata stems from the research into
the structure of quantum grammars and quantum languages but being
simple quantum systems they are natural candidates for elementary
units in programmable quantum systems. From the point of view of
possible applications the theory of quantum lattice gas automata
\cite{may1, may2} deserves special attention. Vlasov \cite{Vlasov}
has considered a simple model of a quantum system whose Hilbert
space $H$ can be decomposed into two components $H=H_{l}\otimes
H_{S}$ where $H_{l} $ corresponds to spatial degrees of freedom of
a hypercubic lattice and $H_{S}$ to internal states. He calls it
the quantum bot (qubot).  The evolution of qubot is described by
conditional quantum dynamics \cite{cond-qd} and its excitation can
be programmed. So far we have considered quantum systems that
occupy a definite space, say a physical laboratory. Benioff
\cite{robot} has considered quantum computers to be parts of
larger systems where interactions between quantum computers and
external systems form an essential part of the overall system
dynamics -- {\em quantum robots}.  A quantum robot is a mobile
system that carries a programmable quantum system,  and all
necessary ancillary systems (e.g. memory) on board. Quantum robots
can carry out tasks whose goals include specified changes in the
state of the environment or carrying out measurements on the
environment. Each task is a sequence of alternating computation
and action phases. Computation phase activities include
determination of the action to be carried out in the next phase
and possible recording of information on neighborhood
environmental system states. Action phase activities include
motion of the quantum robot and changes of neighborhood
environment system states. At this stage  quantum feedback control
seems to be the most effective strategy: we obtain information
about the evolving system through measurement, process the
information and feed it back to the system to actively control the
system in a desired way. Various methods of quantum feedback have
been proposed \cite{feedback}. In Benioff's model each task is
represented by a unitary step operator $T$ that describes single
time steps of quantum robot's dynamics. $T = T_{a}+T_{c}$ is a sum
of action phase ($T_{a}$) and computation phase ($T_{c}$) step
operators. Schematic description the task in terms of decision
trees is possible. No definite times or durations are associated
with the phase steps in the tree. Detailed description of a robot
that performs Grover’s search algorithm  is presented in Ref.
\cite{robot}. He has conjectured that there is an equivalent
Church-Turing hypothesis for the collection of all tasks that can
be carried out by quantum robots. It follows that there may be a
similar hypothesis for the class of feasible physical experiments.
\section{Quantum programming}
In classical computer science high level programming languages
allow to master the more and more complex hardware. Currently no
quantum computer is available and we have only vague idea what
quantum programming should be like. Classical concepts like
hardware abstraction, data classification, memory management, can
hierarchical and structured programming should have quantum
counterparts. The purpose of programming languages is both to
express the semantics of the computation and generation sequences
of elementary operations to be performed by a concrete computing
unit. From this point of view the formalism of Hilbert spaces and
their transformation as the mathematical description of quantum
algorithms provides no means to derive their representation as
sequences of elementary operations to control a given quantum
hardware. Currently known quantum algorithms are described in
terms of {\em quantum random access machine} (QRAM) model
\cite{K-N} that is an extension of a classical  random access
machine which is capable  of both quantum and classical
computations. In such machines the master classical machine uses
the quantum subsystems as a black-box or oracle co-processing
unit. The no-cloning theorem excludes replications of quantum
systems. To handle this problem a new type of data, the {\em
quantum register} has been introduced. Quantum register objects
are collections of  cubits addresses. They can overlap. Quantum
operators  encode definitions of quantum circuits and execute the
circuits on supplied registers. Sanders and Zuliani \cite{S-Z}
extended probabilistic versions of imperative languages to include
quantum primitives. The resulting language ({\em qGCL}) is capable
of programming universal quantum computer. \"{O}mer has given an
excellent analysis of quantum structured programming and developed
a procedural formalism called {\em QCL} in his PhD thesis
\cite{omer}. In both languages unitary transformations are
functions and their manipulation is difficult. To solve this
inconvenience,  Bettelli, Calarco and Serafini \cite{BCS} put
forward an architecture that is capable of compact expression and
reduction to sequences of elementary operation of quantum
algorithms due to introduction of {\em quantum operator objects}
that are easy to handle. Altenkirch and Grattage have taken more
abstract path and introduced a functional language for quantum
computation of finite types - the language QML \cite{QML}. The
programs are interpreted by morphism in the category {\bf FQC} of
finite quantum computation\footnote{ The classical counterpart
{\bf FCC} is also introduced}. Objects of these categories are
sets. Classical computations are carried out on elements of finite
sets and quantum computations take place in finite dimensional
Hilbert spaces. A reversible finite computation is modelled by a
reversible operation, which is a bijection of finite set in the
clasical case, and a unitary transformation on the Hilbert space
in the quantum case. Guided by this they have described the
semantics the language {\em QML} that extends a classical finitary
language. They are currently working on implementation of a
compiler for QML in Haskell. Game theory in the form of
competitive analysis is now a well established tool in analysis of
algorithms \cite{bey}. Quantum games being on the verge of various
approaches seem to be essentially more dynamical than the
traditional games in the ``gaming situations'' that arise from
computational problems \cite{abram, PS1, czachor}. Semantic
analysis of these games reformulated as Hilbert space problem and
the categorical technics should set off differences and
similarities between classical and quantum descriptions. The
formalization must support strategies that are sensitive to any
aspects of the situation, including not just the opponents' moves,
but also the assumptions about their counter-strategies that can
make the most of quantum phenomena such as interaction-free
measurements or counterfactual computations. The hope is that the
careful analysis will provide new defence protocols -- sustainable
strategies against possible attacks\footnote{ and vice versa,
unfortunately.}.  The quantum description would support systematic
development and provide means for dealing with the future
challenges and complexities of real systems \cite{NC}.

\section{Quantum artificial intelligence}
Analogously to the terminology used in computer science, we can
distinguish the shell (the measuring part) and the kernel (the
part being measured) in a quantum game that is perceived as an
algorithm  implemented by a specific quantum process. Note that
this distinction was introduced on the basis of abstract
properties of the game (quantum algorithm, quantum software) and
not properties of the specific physical implementation. Quantum
hardware would certainly require a lot of additional measurements
that are nor specific to the game (or software), cf. the process
of starting a one-way quantum computer. Adherents of artificial
intelligence (AI) should welcome the great number of new
possibilities offered by quantum approach to AI (QAI). For
example, consider a Quantum Game Model of Mind (QGMM) exploring
the confrontation of quantum dichotomy between kernel and shell
with the principal assumption of psychoanalysis of dichotomy
between consciousness and unconsciousness \cite{freud}. The
relation is as follows.
\begin{itemize}
\item Kernel represents the Ego, that is the conscious or more
precisely, that level of the psyche the it is aware of its
existence (it is measured by the Id).  This level is  measured due
to its coupling to the Id via the actual or latent (not yet
measured) carriers of consciousness (in our case qubits
representing strategies) \item Shell represents the Id that is not
self-conscious. Its task is monitoring (that is measuring) the
kernel. Memes, the AI viruses \cite{memy}, can be nesting in that
part of the psyche.
\end{itemize}
Memes being qutrojans, that is quantum parasitic gates  (not
qubits!) can replicate themselves (qubits cannot -- no-cloning
theorem). Very little is known about the possible threat posed by
qutrojans to the future of quantum networks. In quantum
cryptography teleportation of qubits will be helpful in overcoming
potential threats posed by qutrojans therefore we should only
worry about attacks by conventional trojans \cite{lo99}. If the
qutrojan is able replicate itself, it certainly deserves the name
quvirus.  A consistent quantum mechanism of such replication is
especially welcome if quantum computers and cryptography are to
become a successful technology. In the QGMM approach external
measuring apparatus and ``bombs'' reducing (projecting) quantum
states of the game play the role of the nervous system providing
the ``organism'' with contact with the environment that sets the
rules of the game defined in terms of supplies and admissible
methods of using of tactics and pay-offs \cite{komp}. Contrary to
the  quantum automaton put forward by Albert \cite{albert} in QGMM
model there is no self-consciousness -- only the Ego is conscious
(partially) via alliances with the Id and is infallible only if
the Id is not infected with memes. Alliances between the kernel
and the Id (shell) form states of consciousness of QAI and can be
neutralized (suppressed) in a way analogous to the quantum
solution to the Newcomb's paradox \cite{PSN}. In the context of
unique properties of quantum algorithms and their potential
applications the problem of deciding which model of AI (if any)
faithfully describes human mind is fascinating but a secondary
one. The discussed above variant of the Elitzur-Vaidman breaker
suggests that the addition of the third qubit to the kernel could
be useful in modelling the process of forming the psyche by
successive decoupling qubits from the direct measurement domain
(and thus becoming independent of the shell functions). For
example dreams and hypnosis could take place in  shell domains
that are temporary coupled to the kernel in this way. The example
discussed in the previous section illustrates what QAI intuition
resulting in a classically unconveyable  belief might be like.
What important is,  QAI reveals more subtle properties than its
classical counterparts because it can deal with counterfactual
situations \cite{ vai, mitchison} and in that sense analyze
hypothetical situations (imagination). Therefore QAI is
anti-Jourdainian: Molier's Jourdain speaks in prose without
knowing it; QAI might be unable to speak but know it would have
spoken in prose  were it possible. The idea of {\em strong
artificial life} of  building (computational) models  that are so
life-like that they cease to be models of life and become examples
of life themselves should also be invoked here
\cite{Langton,artlife}. An agent based model consists of a
collection of "primitive" computational entities\footnote{For
example, a cellular automata}, called agents. Such agents can in
principle be implemented in the form of quantum programable
systems. Quantum game theoretical aspects of such models have not
yet been investigated. In their particular form known as {\em
artificial chemistry} \cite{artchem} genuinely novel phenomena may
arise at the level of collective interactions of quantum agents
\cite{iqb-dr}. It certainly would influence the discussion of
classical paradigms of artificial intelligence \cite{sul3}.

\section{Conclusion}
Classical computing, though successful, is certainly not the full
story. The opinion that it encompasses only a subset all
computational possibilities has growing number of supporters.
 Humans have already overcome several natural
limitations with help of artificial tools. Are we at the verge of
dramatic developments that would change our computational
paradigms? Quantum information processing with possible
inspirations from physics and biology holds great promise.
Intellectual investment over many years is turning craft into
science. Examining how Nature solves its computational problems
would probably result in  revolutionary changes in computational
paradigms \cite{journeys}. Currently, we  have to accept the
following facts:
\begin{description}
\item The particular choice of physical implementation of
computing units matters and may have consequences \item It may not
be necessary to run the computer to get a result \item The
trajectory taken by a computational process  can be more
interesting than the final result
\end{description}
Quantum game theory also has its weak points but there is no doubt
that it will be a crucial discipline for the emerging information
society. Information processing is undergoing a revolutionary
stage of development. If you ask about its future you get nearly
as many different answers as the number of scientists being asked.
We have presented our personal view that might not come true.
 \begin{center}
{\bf Acknowledgements}
\end{center}
 This paper has been supported
by the {\bf Polish Ministry of Scientific Research and Information
Technology} under the (solicited) grant No {\bf
PBZ-MIN-008/P03/2003}.

\end{document}